\documentclass[prb,twocolumn,showpacs,amsmath,amssymb,aps]{revtex4}

\usepackage{graphicx}
\usepackage{dcolumn}
\usepackage{bm}

\begin{document}


\title{Nonlocal Spin Transport in Lateral Spin Valves with Multiple Ferromagnetic Electrodes}
\author{Tae-Suk Kim and Hyun-Woo Lee}
\affiliation{Department of Physics, Pohang University of Science and Technology,
    Pohang 790-784, Korea}
\author{B. C. Lee}
\affiliation{Department of Physics, Inha University, Incheon 402-751, Korea}

\date{\today}

\begin{abstract}
 We study the nonlocal spin transport in a lateral spin valve with multiple ferromagnetic (FM) electrodes.
When two current-injecting and two spin current-detecting electrodes are all ferromagnetic,
the number of possible nonlocal spin signal states is four at maximum.
In reality, this number is reduced, depending on the inter-probe distance and the relative magnitudes of 
the spin resistances. Our theoretical results are in agreement with recent experiments of spin injection 
into an Al island, a carbon nanotube, and graphene.

\end{abstract}
\maketitle


\section{Introduction}
 Spin injection \cite{johnson, son} from a ferromagnetic (FM) metal into nonmagnetic (NM) materials 
(metal, semiconductor, insulator)
is very important for device applications and for academic interest. 
The operation of the spin valve, a hybrid structure of FM metal/NM/FM metal, depends on efficient 
spin injection from one FM electrode into the NM layer and spin detection in the other FM electrode.
Typical examples of a spin valve include giant magnetoresistance (GMR) devices \cite{gmr},
magnetic tunnel junctions (MTJ)  \cite{mtj1, mtj2, mtj3},  FM/NM/FM nanopillars \cite{nanopillar}, etc.
The magnetoresistance has been enhanced up to a few hundred percent in the recent MgO-based MTJs
\cite{mgo1,mgo2} and  GMR and MTJ devices are already in commercial markets for magnetic read heads.
MgO-based MTJs are now being used for magnetic random access memory devices. 
However, these {\it vertical} spin valves have some difficulties in integrating them 
into semiconductor electronics.

 In order to integrate spintronic devices into semiconductor electronics, 
{\it lateral} spin valves are more desirable for, e.g., multi terminal devices. 
Due to the increased distance between terminals in lateral spin valves, the spin signal is more
suppressed compared to vertical spin valves. 
Furthermore, spin injection and detection experiments in the two terminal geometry of lateral spin valves 
are complicated and obscured by other effects like the anisotropic MR, the anomalous Hall effect, etc. 
In order to avoid these undesirable effects, a nonlocal spin valve geometry \cite{johnson} in which
the spin current path is spatially separated from the charge current path, was adopted, and 
spin injection and detection were clearly demonstrated in Al wires \cite{jedema}.
  The nonlocal spin injection technique was also used to observe \cite{tinkham, otani}
the (inverse) spin Hall effect in diffusive nonmagnetic metals (e.g., Pt with a spin diffusion length of 10 nm), 
which are characterized by a rather strong spin-orbit interaction.
In these experiments, the separation of charge and spin currents, as well as efficient spin injection,
is essential to observing the Hall voltage induced by the spin current.

\begin{figure}[b!]
\includegraphics[width=7.0cm]{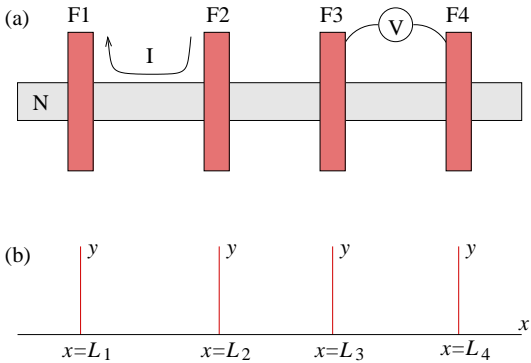}
\caption{Schematic display of the lateral spin valve system with four ferromagnetic electrodes. 
(a) Ferromagnetic electrodes are labeled as F1, F2, F3, and F4 from left to right. 
The base electrode is denoted as N. The current $I$ is injected from F2 into N and drained to F1. 
The voltage induced by the spin current is measured between F3 and F4 electrodes. 
(b) The one-dimensional model geometry of the spin valve system in (a).
\label{4fmlead}}
\end{figure}

  If significant magnetoresistance (MR) or nonlocal spin signals are to be achieved in lateral spin valves, 
the spin polarized current should be able to pass through the intervening nonmagnetic layer without losing or 
degrading too much of its spin polarization in the NM layer  or at the FM-NM interface.
Nonmagnetic materials with long spin diffusion lengths (SDLs) are most desirable and are required 
for successful operation of lateral spin valves. 
In this respect, recent experiments of spin injection into carbon systems, carbon nanotube \cite{wees_cnt} 
and graphene \cite{wees_graphene, fuhrer}, are very intriguing. 
A large spin polarization and a large SDL are observed in carbon systems.
In these experiments, the two voltage probes and thr two current probes were all ferromagnetic
in contrast to previous experiments. 
Motivated by these experiments, we study in this work the spin transport in a nonlocal spin valve geometry
with four FM electrodes, as shown schematically in Fig.~\ref{4fmlead}, 
based on the one-dimensional spin drift-diffusion (SDD) equations.

\section{Formalism: spin drift-diffusion equation}
 Spin transport in spin valves can be understood theoretically based on the spin drift-diffusion (SDD) 
equations \cite{johnson, son},  which are a reduced version of the spin-dependent Boltzmann equation.
The SDD equations were shown to be valid when the mean free path is much less than \cite{valet_fert}
or comparable to \cite{penn_stiles} the spin diffusion length. 
The SDL is the length scale over which electrons can preserve their spin information.
The finite SDL in the samples is caused by the spin-flip scattering due to the spin-orbit interaction, 
magnetic impurities, etc.
Though the SDD equations are phenomenological, they have been very successful in explaining the main features
of experiments qualitatively.

 The SDD equations are written down for the spin-dependent electrochemical potential $\mu_{\alpha}$
and an electric current density $j_{\alpha}$.
Here, $\alpha=\pm$ represents the spin-up ($+$) and the spin-down ($-$) states, respectively.
In the one-dimensional device structure, the SDD equations can be written down in a matrix form as
\begin{eqnarray}
\frac{d^2}{dx^2} \begin{pmatrix} \mu_{+} \cr \mu_{-} \end{pmatrix}
  &=& \begin{pmatrix} \frac{1}{ D_{+} \tau_{+-} } & - \frac{1}{ D_{+} \tau_{+-} } \cr
               - \frac{1}{ D_{-} \tau_{-+} } & \frac{1}{ D_{-} \tau_{-+} }
      \end{pmatrix}
    \begin{pmatrix} \mu_{+} \cr \mu_{-} \end{pmatrix},  \\
j_{\alpha} &=& \frac{\sigma_{\alpha}}{e} \frac{d\mu_{\alpha}}{dx}.
\end{eqnarray}
Here ,$D_{\alpha}$ is the diffusion constant for spin direction $\alpha=\pm$, and $\tau_{+-}$ is
the average spin-flip time for an electron from the spin direction $+$ to $-$.
$\sigma_{\alpha}$ is the conductivity for electrons with spin $\alpha$, and $e$ is the absolute value
of the electron charge.
Analyzing the eigenvalues and the eigenvectors of the matrix in the SDD equation, we can find the general
solution of the SDD equations for the electrochemical potential and the corresponding current density \cite{selman}:
\begin{eqnarray}
\frac{1}{je} \begin{pmatrix} \mu_{+} \cr \mu_{-} \end{pmatrix}
  &=& \left[ \frac{x}{\sigma} + A \right]  ~ \begin{pmatrix} 1 \cr 1 \end{pmatrix}
    + \frac{B\lambda}{2} ~e^{-|x|/\lambda} ~
       \begin{pmatrix} \sigma_{+}^{-1} \cr - \sigma_{-}^{-1} \end{pmatrix},   \\
\frac{1}{j} \begin{pmatrix} j_{+} \cr j_{-} \end{pmatrix}
  &=& \frac{1}{\sigma} ~ \begin{pmatrix} \sigma_{+} \cr \sigma_{-} \end{pmatrix}
    \pm \frac{B}{2} ~e^{-|x|/\lambda} ~
       \begin{pmatrix} 1 \cr -1 \end{pmatrix}.
\end{eqnarray}
Here, $\sigma = \sigma_{+} + \sigma_{-}$ is the total conductivity,
and $A$ and $B$ are parameters to be determined by the boundary conditions.
The total or charge current density is constant and uniform in space: $j = j_{+} + j_{-}$.
The spin diffusion length $\lambda$ is defined by the expression
\begin{eqnarray}
\frac{1}{\lambda^{2} }
  &=&  \frac{1}{ D_{+} \tau_{+-} } + \frac{1}{ D_{-} \tau_{-+} }.
\end{eqnarray}
In a one-dimensional device structure, it is more convenient in algebra to use the current instead of
its density. As we shall show below, the use of properly defined material parameters highly simplifies
the algebra.

 We consider the model spin valve structure in Fig.~\ref{4fmlead}. 
The spin-polarized current $I$ flows from F2 into N and F1. 
The spin-dependent electrochemical potential ($\mu_{i\pm}$) and current ($I_{i\pm}$) in each FM electrode 
can be parameterized as
\begin{eqnarray}
\frac{1}{e} \begin{pmatrix} \mu_{i+} \cr \mu_{i-} \end{pmatrix}
  &=& \left[ \frac{I}{\sigma_1 A_1} y ~\delta_{i,1} - \frac{I}{\sigma_2 A_2} y ~\delta_{i,2} + U_i \right] 
      \begin{pmatrix} 1 \cr 1 \end{pmatrix}   \nonumber\\
  && - I_i {\cal R}_i e^{-y/\lambda_i} \begin{pmatrix} [1 + \beta_i]^{-1} \cr - [1 - \beta_i]^{-1} \end{pmatrix}, \\
\begin{pmatrix} I_{i+} \cr I_{i-} \end{pmatrix} 
  &=& \frac{I}{2} \begin{pmatrix} 1 + \beta_1 \cr 1 - \beta_1 \end{pmatrix} \delta_{i,1} 
      - \frac{I}{2} \begin{pmatrix} 1 + \beta_2 \cr 1 - \beta_2 \end{pmatrix} \delta_{i,2}  \nonumber\\
  && + \frac{I_i}{2} e^{-y/\lambda_i} \begin{pmatrix} 1 \cr -1 \end{pmatrix}.
\end{eqnarray}
The index $i$ runs from 1 through 4, labeling the four FM electrodes. 
As will be clear below, $U_i = V_i + \Delta \delta_{i,1}$ and the shift $\Delta$ is introduced to take into
account the current flow in the common electrode. That is, the electrochemical potential in F1 is shifted up
with respect to that in F2, F3, and F4.  
$V_i$ is the voltage drops in each FM electrode F$i$, which is induced by the spin accumulation and diffusion.
$I_i$ measures the spin current leaking into F$i$ and should be determined by the Kirchoff rules at the junctions.
The spin-dependent current is determined by the equation
\begin{eqnarray}
I_{i\alpha} &=& A_i \frac{\sigma_{i\alpha}}{e} \frac{d}{dy} \mu_{i\alpha}.
\end{eqnarray}
Note that the $i$-th FM lead is contacted to the common electrode at $x = L_i$. $A_i$, $\lambda_i$,
$\sigma_i$, and $\beta_i$ are the cross-sectional area, the spin diffusion length, the conductivity,
and the bulk spin polarization in the conductivity of the $i$-th FM lead, respectively. ${\cal R}_i$ is the resistance
of the FM lead over the spin diffusion length and is defined by the relation
\begin{eqnarray}
{\cal R}_i &=& \frac{\lambda_i}{\sigma_i A_i}.
\end{eqnarray}
The spin-dependent conductivity $\sigma_{i\pm}$ can be written in terms of the spin polarization $\beta_i$ as
\begin{eqnarray}
\sigma_{i\pm} &=& \frac{1}{2} \sigma_i (1 \pm \beta_i).
\end{eqnarray}
In the common base electrode, the spin-dependent electrochemical potential ($\mu_{\pm}$) and current ($I_{\pm}$) 
can be written as 
\begin{eqnarray}
\frac{1}{e} \begin{pmatrix} \mu_{+} \cr \mu_{-} \end{pmatrix} 
  &=& - \sum_i J_i {\cal R} e^{-|x - L_i|/\lambda} \begin{pmatrix} 1 \cr -1 \end{pmatrix}  \nonumber\\
  && - \frac{I}{\sigma A} (x - L_2) \begin{pmatrix} 1 \cr 1 \end{pmatrix} \theta (x - L_1) \theta(L_2 - x) \nonumber\\
  && + \frac{I}{\sigma A} (L_2 - L_1) \begin{pmatrix} 1 \cr 1 \end{pmatrix} \theta (L_1 - x),  \\
\begin{pmatrix} I_{+} \cr I_{-} \end{pmatrix} 
  &=& - \frac{I}{2} \begin{pmatrix} 1 \cr 1 \end{pmatrix} \theta (x - L_1) \theta(L_2 - x)   \nonumber\\
  && + \frac{1}{2} \sum_i J_i \mbox{sgn} (x - L_i) ~ e^{-|x-L_i|/\lambda} 
      \begin{pmatrix} 1 \cr -1 \end{pmatrix}.
\end{eqnarray}
Note that the electrochemical potential shift, $\Delta$, is defined by the relation
\begin{eqnarray}
\Delta &=& \frac{I}{\sigma A} (L_2 - L_1).
\end{eqnarray}
Due to the current flow between F1 and F2, the electrochemical potential is shifted up in the region 
$x \leq L_1$ with respect to the region $x \geq L_2$. 
The spin-dependent current is determined by the equation
\begin{eqnarray}
I_{\alpha} &=& A \frac{\sigma_{\alpha}}{e} \frac{d}{dx} \mu_{\alpha}, ~~~~~ \sigma_{\alpha} ~=~ \frac{\sigma}{2}.
\end{eqnarray}
$A$, $\lambda$, and $\sigma$ are the cross-sectional area, the spin diffusion length, and the conductivity
 of the base electrode N, respectively.
${\cal R} = \lambda/\sigma A$ is the resistance of the base electrode, which is defined over the spin diffusion 
length of the base electrode.

There are three sets of unknown parameters, $V_i, I_i$, and $J_i$'s, which we are going to determine  
by using the boundary conditions or the Kirchoff rules at the junctions. 
The above expressions of the electrochemical potential are constructed such that the charge current 
at each junction is conserved and flows from the F2 lead into the common base electrode and finally into the F1 lead. 
Though the spin-flip scattering in the bulk is taken into account, any possible spin-flip scattering at the 
interface between the FM leads and the nonmagnetic base electrode is neglected. 
In this case, the spin current ($I_s = I_{+} - I_{-}$) is also conserved at each junction
and leads to the following constraint:
\begin{eqnarray}
J_i &=& - \frac{1}{2} I_i - \frac{1}{2} \beta_1 I ~ \delta_{i,1} + \frac{1}{2} \beta_2 I ~ \delta_{i,2}.
\end{eqnarray}
The electrochemical potential for each spin direction should satisfy Ohm's law at the junctions.
At the $i$-th junction, the drop in the electrochemical potential is given by the expression
\begin{eqnarray}
\frac{1}{e} \Delta \mu_{i\pm} 
  &=& U_i \mp \frac{{\cal R}_i}{1 \pm \beta_i} I_i   
   - \left[ \mp {\cal R} \sum_{ij} A_{ij} J_j + \Delta ~ \delta_{i,1}  \right]  \nonumber\\
  &=& V_i \mp \frac{{\cal R}_i}{1 \pm \beta_i} I_i   
    \pm {\cal R} \sum_{ij} A_{ij} J_j,  \nonumber  \\
A_{ij} &\equiv& e^{-|L_i - L_j|/\lambda}.
\end{eqnarray}
Ohm's law at the junctions gives the following relation:
\begin{eqnarray}
\frac{1}{e} \Delta \mu_{i\pm} 
  &=& {\cal R}_{ti\pm} I_{i\pm},  
\end{eqnarray}
where ${\cal R}_{ti\pm}$ is the spin-dependent junction resistance at the interface between the base electrode and 
the $i$-th FM electrode F$i$, and can be defined in terms of the spin polarization $\gamma_i$ 
of the junction resistance as
\begin{eqnarray}
{\cal R}_{ti\pm}  
  &=& \frac{2{\cal R}_{ti}}{1 \pm \gamma_i}.
\end{eqnarray}
$I_{i\pm}$ is the spin-dependent current passing through the $i$-th junction:
\begin{eqnarray}
I_{i\pm} 
  &=& \frac{I}{2} (1 \pm \beta_1) \delta_{i,1} - \frac{I}{2} (1 \pm \beta_2) \delta_{i,2} \pm \frac{I_i}{2}. 
\end{eqnarray}  
In our work, two types of spin polarization are introduced. One is $\beta_i$, measuring the spin polarization 
in the bulk conductivity in the FM electrode F$i$. 
The other one is $\gamma_i$, which measures the spin polarization in the spin-dependent junction resistance 
between N and F$i$. 
In our convention, $\beta$ and $\gamma$ are positive when spin-up electrons are in the majority band
while they are negative when the magnetization orientation is reversed or the spin-up electrons are
in the minority band.

For the algebraic manipulation, it is more convenient to introduce new parameters for the resistance:
\begin{eqnarray}
R_i &=& \frac{{\cal R}_i}{1 - \beta_i^2}, ~~~ R_{ti} ~=~ \frac{{\cal R}_{ti}}{1 - \gamma_i^2}, 
\end{eqnarray}
and $R = {\cal R}$. With these new notations, we have 
\begin{eqnarray}
V_i &=& \pm (1 \mp \beta_i) I_i R_i \mp R \sum_{j} A_{ij} J_j  \nonumber\\
    && + R_{ti} (1 \mp \gamma_i) \left[ I (1 \pm \beta_1) \delta_{i,1} 
                                       - I (1 \pm \beta_2) \delta_{i,2} \pm I_i \right].
\end{eqnarray}
Adding and subtracting the two equations, we find
\begin{eqnarray}
V_i &=& - ( \gamma_i R_{ti} + \beta_i R_i ) I_i 
      + (1-\beta_1\gamma_1) R_{t1} I ~\delta_{i,1}  \nonumber\\
    && - (1-\beta_2\gamma_2) R_{t2} I ~\delta_{i,2}, 
\end{eqnarray}
and
\begin{eqnarray}
&& (R_i + R_{ti}) I_i - R \sum_{j} A_{ij} J_j  \nonumber\\
  && \hspace{0.5cm} =  -(\beta_1 - \gamma_1) R_{t1} I ~ \delta_{i,1} 
    +(\beta_2 - \gamma_2) R_{t2} I ~ \delta_{i,2}.
\end{eqnarray}
For the algebraic manipulation, it is much more convenient to use the vector and matrix notations to rewrite 
the equations obtained from the boundary conditions as
\begin{eqnarray}
|V> &=& - ( \hat{\gamma} {\bf R}_t + \hat{\beta} {\bf R} ) |I>   
       + (1-\beta_1\gamma_1) R_{t1} I ~|1>   \nonumber\\
    && - (1-\beta_2\gamma_2) R_{t2} I ~|2>,  \\
|I> &=& - 2|J> - \beta_1 I |1> + \beta_2 I |2>,  
\end{eqnarray}
and
\begin{eqnarray}
&& ({\bf R} + {\bf R}_t ) |I> - R {\bf A} |J>  \nonumber\\
  && \hspace{0.5cm} = -(\beta_1 - \gamma_1) R_{t1} I |1> + (\beta_2 - \gamma_2) R_{t2} I |2>.
\end{eqnarray}
${\bf R}$ and ${\bf R}_t$ are diagonal matrices with their diagonal elements $R_{ii} = R_i$ and
$R_{tii} = R_{ti}$, respectively. We used the vector notations, for example, $<I| = (I_1, I_2, I_3, I_4)$ 
and $<1| = (1,0,0,0)$.  
After some algebra, we find the expression of the desired electrochemical potential drops:
\begin{eqnarray}
\label{V_spin}
\frac{V_i}{I}
  &=& (-1)^{i+1} (R_{ti} + \beta_i^2 R_i) \delta_{i\leq 2}  \nonumber\\
  && + \sum_{j=1}^{2} (\gamma_i R_{ti} + \beta_i R_i) G_{ij} (-1)^{j} (\gamma_j R_{tj} + \beta_j R_j),
\end{eqnarray}
where the matrix ${\bf G}$ is defined by the relation
\begin{eqnarray}
{\bf G} &=& \left[ {\bf R} + {\bf R}_t + \frac{1}{2} R {\bf A} \right]^{-1}.
\end{eqnarray}
Note that the matrix ${\bf G}$ has the dimension of conductance and is independent of the magnetization configurations
in the FM electrodes. 
The spin current $I_i$ is also obtained as
\begin{eqnarray}
\label{I_spin}
\frac{I_i}{I}
  &=& - \beta_1 \delta_{i1} + \beta_2 \delta_{i2} 
     + G_{i1} (\beta_1 R_1 + \gamma_1 R_{t1})  \nonumber\\
  && - G_{i2} (\beta_2 R_2 + \gamma_2 R_{t2}). 
\end{eqnarray} 
$I_3$ and $I_4$ measure the leakage spin currents into F3 and F4, respectively.
Since $G_{ij}$ is independent of the magnetization configurations, the signs and the magnitudes of 
the leaking spin currents do not depend on the magnetization orientations of the F3 and the F4 electrodes
(parallel or antiparallel relative to F1 and F2), but depend on those of F1 and F2.
On the other hand, the voltage drops $V_3$ and $V_4$ change their sign when their magnetizations
are reversed.

\section{Results and discussion}
 In the previous section, we derived the voltage drops and  leaking spin currents 
in the voltage probes F3 and F4. We can consider several different measurement geometries
from our analytic solutions.

 For the voltage probes F3 and F4, we have the spin leakage currents and the corresponding voltage drops 
from Eqs.~(\ref{V_spin}) and (\ref{I_spin}):
\begin{eqnarray}
\frac{I_i}{I}
  &=& G_{i1} (\beta_1 R_1 + \gamma_1 R_{t1}) - G_{i2} (\beta_2 R_2 + \gamma_2 R_{t2}),  \\
V_i &=& - (\gamma_i R_{ti} + \beta_i R_i) I_i.
\end{eqnarray}
Note that the nonlocal voltage drop can be expressed as the product of the leakage spin current 
and the appropriately defined spin resistance. This spin resistance depends on the orientation of the
magnetization. Our general results can be reduced to the simpler ones.
When F1 is located very far away from F2 or when the distance between F1 and F2 or $|L_2-L1|$ is
much longer than the SDL of the nonmagnetic electrode, our spin valve structure is reduced to 
that studied by the authors in Ref.~\onlinecite{KLL}, where the effect of an additional FM electrode
on the nonlocal spin signals was studied.
Furthermore, if the distance between F3 and F4 is much longer than the spin diffusion length,
our model spin valve is reduced to the {\it conventional} spin valve studied in Ref.~\onlinecite{jedema}.

 The nonlocal spin signal in the model spin valve in Fig.~\ref{4fmlead} is quantified 
by measuring the voltage difference between two FM electrodes, F3 and F4. 
Hence $\Delta V = V_3 - V_4$ is the experimentally relevant quantity 
and the nonlocal spin signal is defined as $R_s = \Delta V /I$, where
\begin{eqnarray}
\label{S_signal}
R_s &=& \sum_{i=3,4} \sum_{j=1,2} (-1)^{i+j+1} (\gamma_i R_{ti} + \beta_i R_i) G_{ij}  \nonumber\\
    &&  \times (\gamma_j R_{tj} + \beta_j R_j).
\end{eqnarray} 
This is the main result of our work.

 To get some insight about the nonlocal spin signal $R_s$,
let us start with an algebraic manipulation of the conductance matrix ${\bf G}$.
For the calculation of the inverse matrix $G$, we note the symmetry of $G$ and $G^{-1}$ can be rewritten as
\begin{eqnarray}
G^{-1} &=& \begin{pmatrix} A & T \cr T^{\dag} & B \end{pmatrix},  \\
A &=& \begin{pmatrix} r_1 & \frac{1}{2} f_1 R \cr \frac{1}{2} f_1 R & r_2 \end{pmatrix},  \\
B &=& \begin{pmatrix} r_3 & \frac{1}{2} f_3 R \cr \frac{1}{2} f_3 R & r_4 \end{pmatrix},  \\
T &=& \frac{1}{2} f_2 R ~ \begin{pmatrix} f_1 & f_1 f_3 \cr 1 & f_3 \end{pmatrix}.
\end{eqnarray}
Here, $f_i = e^{-|L_{i+1} - L_i|/\lambda}$ with $i=1,2,3$, and $r_i = R_i + R_{ti} + \frac{1}{2}R$
with $i=1,2,3,4$. 
The $2\times 2$ matrix $T$ is real, and its hermitian $T^{\dag}$ is equivalent to its transpose $T^{t}$.
Though $|T|=0$, the determinants of $A$ and $B$ are nonzero so that we can find the conductance matrix $G$ 
in a manageable form. The inverse of a $4\times 4$ matrix is reduced to that of $2\times 2$ matrices, 
which is much simpler to compute: 
\begin{widetext}
\begin{eqnarray}
G &=& \begin{pmatrix}
      A^{-1}[1 - TB^{-1}T^{\dag}A^{-1}]^{-1} & - A^{-1} T B^{-1} [1 - T^{\dag} A^{-1}TB^{-1}]^{-1}  \cr
     - B^{-1} T^{\dag} A^{-1} [1 - TB^{-1}T^{\dag}A^{-1}]^{-1}  & B^{-1} [1 - T^{\dag} A^{-1}TB^{-1}]^{-1} 
      \end{pmatrix}.
\end{eqnarray}
For the nonlocal spin signals such as $I_3,I_4$ and $V_3, V_4$, only the off-diagonal block matrix is relevant. 
Since $T$ contains the exponentially decaying factor with the scale of the spin diffusion length, we may 
approximate the matrix $G$ in the expression of a nonlocal spin signal as
\begin{eqnarray}
\begin{pmatrix} G_{31} & G_{32} \cr G_{41} & G_{42} \end{pmatrix}
  &=& - B^{-1} T^{\dag} A^{-1} [1 - TB^{-1}T^{\dag}A^{-1}]^{-1} 
  ~\approx~ - B^{-1} T^{\dag} A^{-1}  \nonumber\\
  &=& - \frac{f_2 R}{2 (r_1r_2 - \frac{1}{4}f_1^2 R^2) (r_3r_4 - \frac{1}{4}f_3^2 R^2) }
    \begin{pmatrix} f_1 (r_2 - \frac{R}{2})(r_4 - \frac{1}{2} f_3^2 R) 
                     & (r_1 - \frac{1}{2} f_1^2R)(r_4 - \frac{1}{2} f_3^2R) \cr
                  f_1 f_3 (r_2 - \frac{1}{2} R) (r_3 - \frac{1}{2} R)
                     & f_3(r_1 - \frac{1}{2} f_1^2R)(r_3 - \frac{1}{2} R)
    \end{pmatrix}.
\end{eqnarray}
\end{widetext}
This approximate form is valid when the distance between the current probe F2 and the voltage probe F3 
is much larger than the SDL so that the matrix $T$ is small. 
For a more accurate calculation, the exact form of the conductance matrix ${\bf G}$ is needed. 
However, the above approximate conductance matrix is enough to get some insight into the nonlocal
spin signal $R_s$.

Since $f_i$ is an exponentially decaying factor over the spin diffusion length, the most dominant contribution 
in the nonlocal spin signal $R_s$ comes from $G_{32}$. The nonlocal spin signal or the transresistance $R_s$ 
in Eq.~(\ref{S_signal}) can be rewritten in decreasing order as 
\begin{eqnarray}
R_s  &=& (\gamma_3 R_{t3} + \beta_3 R_3) G_{32} (\gamma_2 R_{t2} + \beta_2 R_2)  \nonumber\\
  && - (\gamma_3 R_{t3} + \beta_3 R_3) G_{31} (\gamma_1 R_{t1} + \beta_1 R_1)  \nonumber\\
  && - (\gamma_4 R_{t4} + \beta_4 R_4) G_{42} (\gamma_2 R_{t2} + \beta_2 R_2)  \nonumber\\
  && + (\gamma_4 R_{t4} + \beta_4 R_4) G_{41} (\gamma_1 R_{t1} + \beta_1 R_1).
\end{eqnarray} 
Due to the exponential factors, the first line is dominant and the last line is the weakest.
Obviously, the nonlocal spin signal will change its sign when the magnetization orientation of
two neighboring FM electrodes F2 and F3 is inverted.

 In experiments, the magnetic configurations are scanned by sweeping the external magnetic field
along the geometrically parallel FM electrodes. With different widths of the FM electrodes, the coercive
fields are different due to the different strengths of an easy axis anisotropy along the electrode direction.
Starting from a large negative field, all the FM electrodes are aligned with the magnetic field.
As the magnetic field is swept from negative to positive, the FM electrodes reverse their
magnetization one by one when the magnetic field equals their coercive field.
While sweeping the magnetic field from large negative to large positive, we have four different 
magnetization configurations so that nonlocal spin signal will show four different states.
When the magnetic field is now swept from large positive to large negative, we have again four
different magnetization configurations.
Under the time reversal operation or when the magnetization orientation of all four FM electrodes 
is reversed, we can readily see that the nonlocal spin signal $R_s$ does not change its sign.
This means that there are only four different values of $R_s$ under the magnetic field sweeping.  
In principle, there can be 4 different states in the nonlocal spin signal, depending on the relative orientations 
of magnetizations. However, the number of actual states realized in experiments may well 
depend on the distance between the FM contacts, the relative magnitudes of the spin resistances in $R_s$.
We note that all four states were observed in Ref.~\onlinecite{wees_Al}  for a nonlocal spin valve 
with Al as a base electrode, three states in Ref.~\onlinecite{wees_graphene}, 
and two states in Ref.~\onlinecite{wees_cnt}.

\section{Summary and conclusion}
 In this paper, we studied the nonlocal spin transport in a lateral spin valve with four FM electrodes,
out of which two are used as current injectors and the other two are used as spin current detectors or 
the voltage probes. We calculated the general expressions for the nonlocal signals, such as the leakage spin current
and the voltage drop due to the spin accumulation. Since there are four ferromagnetic electrodes in 
our model spin valve structure, in principle, four different nonlocal spin signal states are possible.
In real experiments, the number of observed states is found to depend on the inter-electrode
distance, the relative magnitude of spin resistances.

\acknowledgments
This work was supported by a Korea Science and Engineering Foundation (KOSEF) grant funded by
the Korea government (Ministry of Science and Technology (MOST), No. R01-2005-000-10303-0), 
by the SRC/ERC program of MOST/KOSEF (R11-2000-071)
and by the POSTECH Core Research Program.


%
%
%

\end{document}